\documentclass[12pt]{article}

\usepackage{amssymb}
\usepackage[dvips]{graphicx}

\def\appendix#1{
  \addtocounter{section}{1}
  \setcounter{equation}{0}
  \renewcommand{\thesection}{\Alph{section}}
  \section*{Appendix \thesection\protect\indent \parbox[t]{11.715cm} {#1}}
  \addcontentsline{toc}{section}{Appendix \thesection\ \ \ #1}
  }

\newcommand {\bd}{\begin{displaymath}}
\newcommand {\ed}{\end{displaymath}}
\newcommand {\eq}{\begin{equation}}
\newcommand {\beq}{\begin{equation}}
\newcommand {\eeq}{\end{equation}}
\newcommand {\beqa}{\begin{eqnarray}}
\newcommand {\eeqa}{\end{eqnarray}}
\newcommand {\n}{\nonumber \\}
\newcommand {\tr}{{\rm tr\,}}
\newcommand {\Tr}{\mbox{Tr\,}}

\newcommand {\ee}{\mbox{e}}

\newcommand {\defeq}{\stackrel{\rm def}{=}}

\font\mybb=msbm10 at 12pt
\def\bb#1{\hbox{\mybb#1}}

\def\IC{{\bb C}}
\def\IR{{\bb R}}
\def\IZ{{\bb Z}}

\newcommand{\id}{{1\!\!1}} 

\begin{document}

\setlength{\oddsidemargin}{0cm}
\setlength{\baselineskip}{7mm}

\begin{titlepage}

\baselineskip=14pt

 \renewcommand{\thefootnote}{\fnsymbol{footnote}}
\begin{normalsize}
\begin{flushright}
\begin{tabular}{l}
NBI-HE-00-08\\
ITEP-TH-10/00\\
hep-th/0002158\\
\hfill{ }\\
February 2000
\end{tabular}
\end{flushright}
  \end{normalsize}

{}~~\\

\vspace*{0cm}
    \begin{Large}
       \begin{center}
{Nonperturbative Dynamics of Noncommutative Gauge Theory}\\

       \end{center}
    \end{Large}
\vspace{1cm}

\begin{center}
           J. A{\sc mbj\o rn}$^{1)}$\footnote
            {
e-mail address :
ambjorn@nbi.dk},
           Y.M. M{\sc akeenko}$^{1)\,2)}$\footnote
            {
e-mail address :
makeenko@nbi.dk},
           J. N{\sc ishimura}$^{1)}$\footnote{
Permanent address : Department of Physics, Nagoya University,
Nagoya 464-8602, Japan,\\
e-mail address : nisimura@nbi.dk}
           {\sc and}
           R.J. S{\sc zabo}$^{1)}$\footnote
           {e-mail address : szabo@nbi.dk}\\
      \vspace{1cm}
        $^{1)}$ {\it The Niels Bohr Institute\\ Blegdamsvej 17, DK-2100
                 Copenhagen \O, Denmark}\\[4mm]
        $^{2)}$ {\it Institute of Theoretical and Experimental Physics}\\
               {\it B. Cheremushkinskaya 25, 117218 Moscow, Russia} \\
\end{center}

\vskip 1 cm

\hspace{5cm}

\begin{abstract}
\noindent
We present a nonperturbative lattice formulation of noncommutative
Yang-Mills theories in arbitrary even dimension.
We show that lattice regularization
of a noncommutative field theory requires finite lattice volume
which automatically provides both an ultraviolet and an infrared cutoff.
We demonstrate explicitly Morita equivalence of commutative
U($p$) gauge theory with $p\cdot n_f$ flavours of 
fundamental matter fields on a lattice 
of size $L$ with twisted boundary conditions 
and noncommutative U(1) gauge theory with 
$n_f$ species of matter on a lattice 
of size $p\cdot L$ with single-valued fields.
We discuss the relation with twisted large $N$ reduced models and
construct observables in noncommutative gauge theory with matter.
\end{abstract}
\vfill
\end{titlepage}
\vfil\eject
\setcounter{footnote}{0}

\setcounter{equation}{0}
\renewcommand{\thefootnote}{\arabic{footnote}}

\baselineskip=15pt

\setcounter{footnote}{0}
\section{Introduction}
\setcounter{equation}{0}

\renewcommand{\thefootnote}{\arabic{footnote}}
It has been suggested for some time that noncommutative geometry is a natural
framework to describe nonperturbative string theory.
This belief has been supported by the fact that
Matrix Theory \cite{BFSS} or the IIB matrix model \cite{IKKT},
which are conjectured to provide nonperturbative definitions of string
theories, give rise to noncommutative Yang-Mills theory on a toroidal
compactification \cite{CDS}. The particular toroidal compactification
can be interpreted in terms of the presence of a background $B$ field
\cite{bigatti},
which can also be understood in the context of open string quantization
\cite{SW}. Noncommutative gauge theories possess a number of important
properties inherent from noncommutative geometry, where
there is a remarkable geometric equivalence
relation on certain classes of noncommutative spaces known as Morita
equivalence. 
In noncommutative Yang-Mills theory, this implies a duality between
gauge theories over different noncommutative tori, for example, which relates a
Yang-Mills theory with background magnetic flux to a gauge theory with gauge
group of lower rank and no background flux. It allows one to interpolate
continuously, through noncommutative Yang-Mills theories, between two ordinary
Yang-Mills theories with gauge groups of different rank and appropriate
background magnetic fluxes. Furthermore, in certain instances, there is the
remarkable fact~\cite{Morita} 
that the non-abelian nature of a gauge group can be absorbed
into the noncommutativity of spacetime by mapping a U$(p)$ gauge theory with
multi-valued gauge fields to a U(1) gauge theory with single-valued fields on a
dual noncommutative torus.

While most of the results concerning Morita equivalence are obtained
at the classical level, there are strong indications that it
persists in regularized perturbation theory. In particular, the one-loop
divergences coincide~\cite{ren} for noncommutative
 gauge theory and ordinary commutative Yang-Mills theory on $\IR^4$
after a proper rescaling of the coupling constant. Moreover,
it can be shown~\cite{AIIKKT,IIKK,IKK} that noncommutative gauge theory
is equivalent to all orders of perturbation theory to
a twisted large $N$ reduced model~\cite{EK,GO}. 
Following this technique, gauge-invariant observables for
noncommutative Yang-Mills theory have been constructed.
A surprising new class of observables, which 
are expressed in terms of open 
Wilson loops, exists in noncommutative gauge 
theory~\cite{IIKK,AMNS} in addition to those expressed via a counterpart of  
the standard closed Wilson loops of the reduced model.

Lattice gauge theories~\cite{Wilson} are a standard tool for nonperturbative
investigations of nonsupersymmetric gauge theories. In this Letter we will
apply a lattice formulation of noncommutative gauge theory to study its 
nonperturbative properties and, in particular, Morita equivalence. The
construction is an extension of a previous work~\cite{AMNS} in which
a unified framework was developed which naturally interpolates between
the Matrix theory compactification and the twisted large $N$ reduced model
versions of noncommutative Yang-Mills theory.
It was shown that a finite $N$ matrix model obtained from a constrained
twisted Eguchi-Kawai model is equivalent
to a manifestly star-gauge invariant formulation of
noncommutative U(1) gauge theory on a lattice of finite extent.
In the following 
we will further investigate the properties of noncommutative Yang-Mills
theories in arbitrary even dimension, and in particular
reconsider the lattice formulation of noncommutative gauge theory
from a general point of view,
without specifying a particular representation of the noncommutative algebra.
Because the lattice provides a regularization of the theory,
all the results obtained are rigorous in this sense.
We will show that in noncommutative geometry,
when one discretizes spacetime,
consistency of the algebra requires 
 spacetime to be compactified as well, thereby automatically providing
an infrared cutoff.
This is similar to the correspondence between
infrared and ultraviolet in noncommutative field theories
discovered~\cite{uvir} in perturbation theory.
We will also explicitly construct a map between the fields in commutative
U($p$) gauge theory on a lattice 
of size $L$ with twisted boundary conditions
(representing the existence of a 't~Hooft flux) and noncommutative
U(1) gauge theory on a lattice 
of size $p\cdot L$ with periodic boundary conditions,
thereby demonstrating their Morita equivalence.

The technique introduced also allows us to study properties of
noncommutative gauge theory with matter. We will consider the coupling to
matter fields in the fundamental
representation of the gauge group and explicitly demonstrate Morita equivalence
between commutative U($p$) gauge theory with 
$p\cdot n_f$ different flavours of matter on a lattice 
of size $L$ with twisted boundary conditions 
and noncommutative U(1) gauge theory with 
$n_f$ flavours of matter on a lattice 
of size $p\cdot L$ with periodic boundary conditions. 
In the simplest case of $L=1$ we recover the
reduced model introduced in~\cite{Das}.
We discuss observables represented by the matter fields
and rewrite them in terms of Wilson loops by integrating out
the matter. This is the analog of the original construction~\cite{Wilson}
of observables of ordinary lattice gauge theory from matter field averages.
We will construct, along these lines,
observables which correspond to open Wilson
loops and emphasize their role in the nonperturbative dynamics of
noncommutative gauge theory.

\section{Noncommutative lattice gauge theory}
\label{review}
\setcounter{equation}{0}

On a noncommutative space of dimension $D$, the 
 local coordinates $x_\mu$ are replaced by
hermitian operators $\hat{x}_\mu$ obeying the commutation relations
$ [\hat{x}_\mu,\hat{x}_\nu] = i\,\theta_{\mu\nu} $,
where $\theta_{\mu\nu}=-\theta_{\nu\mu}$ are (dimensionful) real-valued
c-numbers. To describe a lattice discretization, we
restrict the spacetime points to
$x_\mu\in\epsilon\IZ$, where $\epsilon$ is the lattice spacing. 
The lattice momentum then has the periodicity $
k _\mu \rightarrow k _\mu + \frac{2\pi}{\epsilon} \delta _{\mu\nu}$,
$\nu = 1,\dots , D$, 
which implies an identity of operators such as
$
\ee^{ i (k_\mu  + {2 \pi}\delta _{\mu\nu}/{\epsilon} ) 
\hat{x}_\mu} 
= \ee^{ i  k_\mu  \hat{x}_\mu}$.
Acting by $\ee^{- i  k_\mu  \hat{x}_\mu}$ on both sides,
we conclude that
$\ee^{2 \pi i \hat{x}_\mu /\epsilon} = \id $
and the momentum $k_\mu$ should be restricted as
\beq
\theta _{\mu\nu} k_\nu\in2 \epsilon  \IZ  
\label{latticemomentum}
\eeq
with $\frac{\pi}{\epsilon ^2}\,\theta_{\mu\nu}$
a $D \times D$ integer-valued matrix. Note that
the restriction (\ref{latticemomentum}) on the momentum 
simply disappears in the commutative case $\theta_{\mu\nu}=0$,
and is quite characteristic of the noncommutative geometry.

The discretization (\ref{latticemomentum})
of the momentum implies that
the lattice is periodic, $ x_\mu \sim x _\mu + 
\Sigma _{\mu\nu}$, $\nu = 1, \dots , D$, with period matrix $\Sigma$ given by
\beq
 M _{\mu \lambda}  \Sigma _{\nu\lambda} = 
\frac{\pi}{\epsilon}\,\theta _{\mu\nu} \ ,
\label{MST}
\eeq
where $M _{\mu\nu}$ is a $D \times D$ integer-valued matrix.
Thus, we have found that lattice regularization of noncommutative field theory
inevitably requires the lattice to be compact as well. 
In the continuum limit
$\epsilon \rightarrow 0$, 
the infrared cutoff goes away as $\frac{1}{\epsilon}$.
For any given periodicity $\Sigma _{\mu\nu}$ and 
noncommutativity $\theta _{\mu\nu}$ in the continuum,
one can construct a series of lattice theories 
satisfying the restriction (\ref{MST}) and
approaching the target continuum theory
in the $\epsilon \rightarrow 0$ limit.
 
Lattice fields $\phi_i (x)$ on the $D$ dimensional noncommutative
spacetime are now replaced by finite dimensional operators $\hat\phi_i$. 
Their coordinate-space representation $\phi_i (x)$ can be obtained
using a map
\beq
\hat{\phi}_i =  \sum_x\hat{\Delta} (x)\, \phi_i (x) 
\label{NCPhi}
\eeq
with
\beq
\hat{\Delta} (x) =
\frac1{|\det \frac1\epsilon\,\Sigma|}\,
\sum_{\vec{m}}
\left(\prod_{\mu=1}^D\left(\hat Z _\mu\right)^{m_\mu}\right)
{}~\ee ^{\pi i\sum_{\mu<\nu}\Theta_{\mu\nu}m_\mu m_\nu}~
\ee ^{- 2 \pi i (\Sigma ^{-1})_{\mu\nu} m_\mu x_\nu }=\hat\Delta(x)^\dag \ ,
\label{map_cont_torus}
\eeq
where the sum in (\ref{NCPhi}) runs over all lattice points modulo the
lattice periodicity, and the sum in (\ref{map_cont_torus}) goes over all
integers $m_\mu$ modulo the periodicity $\frac1\epsilon\,\Sigma_{\nu\mu}$.
The operators 
\beq
\hat Z_\mu = 
\ee ^{2 \pi i (\Sigma ^{-1}) _{\mu\nu} \hat{x}_\nu } \, 
\eeq
satisfy the commutation relations
\beq
\hat Z_\mu\hat Z_\nu=\ee^{- 2\pi i\Theta_{\mu\nu}}\,\hat Z_\nu\hat
Z_\mu\label{Zcomms} \ ,
\label{delZcomms}\eeq
where the dimensionless noncommutativity parameter
\beq
\Theta_{\mu\nu} =
2\pi (\Sigma ^{-1})_{\mu\lambda} \theta_{\lambda\rho} 
(\Sigma ^{-1})_{\nu\rho}
\label{Thetadimless}\eeq
is necessarily rational-valued on the lattice,
since the restriction (\ref{MST}) implies that 
\beq
M_{\mu\nu} = \frac1{2\epsilon} \Sigma_{\mu\lambda} \Theta_{\lambda\nu}
\label{MSTdimless}
\eeq
should be an integer-valued matrix.
The normalization of the trace
of operators is fixed by $\Tr\hat{\Delta} (x)=1$. The
collection of operators $\hat\Delta(x)$ form an orthonormal set and
the inverse of the map (\ref{NCPhi}) is given by
$ \phi_i(x)=\Tr(\hat\phi_i\,\hat\Delta(x))$.
The product
$\hat\phi_1\hat\phi_2$ of two operators has coordinate space representation
given by the lattice star-product
\beq
\Tr\Bigl(\hat\phi_1\hat\phi_2\,\hat\Delta(x)\Bigr)=
\frac{1}{|\det \frac1\epsilon\,\Sigma|}\,
\sum_{y,z} \phi_1(y)\,\phi_2(z)\;
\ee^{- 2 i (\theta^{-1})_{\mu\nu} (x_\mu-y_\mu) (x_\nu-z_\nu)}
\defeq\phi_1(x)\star\phi_2(x)
\label{latticekernelsimple}
\eeq
where we have assumed that $(M ^{-1})_{\mu\nu}$ 
is an integer-valued matrix\footnote{This formula 
is similar to Ref.~\cite{BM}.
See also \cite{earlier} for earlier works in this regard.}.

In order to construct a lattice formulation of noncommutative Yang-Mills
theory, we need to maintain star-gauge invariance on the lattice.
As in the case of ordinary lattice gauge theory \cite{Wilson}, 
this is achieved by putting the U$(n)$ gauge fields 
on the links of the lattice~\cite{AMNS}.
This determines a unitary operator
\beq
\hat{U}_\mu =  \sum _x\hat{\Delta} (x) \otimes U_\mu (x) \ ,
\label{defUhat}
\eeq
where  $U_\mu(x)$
is an $n \times n$ matrix field on the lattice which is star-unitary,
$ U_\mu(x) \star U_\mu(x)^{\dag}=\id_n $.
The lattice action is
\beq
S = -\frac1{g^2}\,
\sum _x\sum_{\mu\neq\nu}\tr_{(n)}\Bigl[U_\mu (x) \star
U_\nu (x + \epsilon \hat{\mu}) \star
U_\mu (x + \epsilon \hat{\nu}) ^\dag \star
U_\nu (x) ^\dag\Bigr] \ ,
\label{latticeaction}
\eeq
and it is invariant under the lattice star-gauge transformation
\beq
U_\mu (x)\mapsto g(x) \star U_\mu (x)
\star g(x+\epsilon \hat{\mu})^\dag  \ ,
\label{latticestargaugetr}
\eeq
where the gauge function $g(x)$  
is star-unitary, $g(x)\star g(x)^{\dag}=\id_n$.
The lattice has finite extent, as is already discussed, and
the fields $U_\mu(x)$ are single-valued,
$U_\mu(x+\Sigma _{\alpha\nu} \hat\alpha) =U_\mu(x)$.

\section{Lattice Morita equivalence}
\label{lattice}
\setcounter{equation}{0}

We will now demonstrate that the noncommutative Yang-Mills theory with 
action~(\ref{latticeaction})
is Morita equivalent to a commutative Yang-Mills theory on
a lattice of different size and with a different gauge group, thereby 
generalizing previous results~\cite{AIIKKT,AMNS}
 for the twisted Eguchi-Kawai model.
We start with a {\it commutative} U$(p)$ lattice gauge theory
with 't~Hooft fluxes.  The action is
\beq
S =-\frac{1}{\tilde{g}^2}\,
\sum _{x}  \sum_{\mu \neq \nu} 
\tr_{(p)}\Bigl[U_\mu (x)\,
U_\nu (x + \epsilon \hat{\mu})\,
U_\mu (x + \epsilon \hat{\nu}) ^\dag\,
U_\nu (x) ^\dag\Bigr] \ ,
\label{commutativeaction}
\eeq
where $U_\mu (x)$ are U($p$)
gauge fields satisfying the twisted boundary conditions
\beq
U_\mu (x+ \Sigma _{\alpha\nu} \hat{\alpha}) =
\Omega _\nu(x)\,U_\mu (x)\,\Omega _\nu(x+ \epsilon \hat{\mu})^\dag \  ,
\label{bc}
\eeq
with period matrix $\Sigma$.
The transition functions $\Omega_\mu (x)$ are SU$(p)$  matrices
obeying the consistency condition~\cite{tHooft}
\beq
\Omega _\mu (x + \Sigma _{\alpha\nu} \hat{\alpha}) ~ \Omega _\nu (x)
= {\cal Z}_{\mu\nu}~ \Omega _\nu (x+ \Sigma _{\alpha\mu} \hat{\alpha}) ~ 
\Omega _\mu (x) \,,
\label{cocycleD}
\eeq 
where ${\cal Z}_{\mu\nu}=\ee^{2\pi i Q_{\mu\nu}/p}\in\IZ_p$.
The antisymmetric matrix $Q$ has elements $Q_{\mu\nu}\in\IZ$ 
representing the 't~Hooft fluxes. The reason why we have chosen
the $\Omega_\mu (x)$ as SU$(p)$, rather than the usual U$(p)$, matrices is
that the abelian magnetic flux of a U$(p)$ gauge field, 
which plays a very important role in Morita equivalences of 
noncommutative gauge theories~\cite{Morita}, arises in the
commutative case only via the 
corresponding 't~Hooft flux~\cite{LPR}.
A similar consideration with U$(p)$ transition matrices
would lead us to the same results.

We take the gauge choice
$ \Omega_\mu(x)= \Gamma_\mu $
with constant $\Gamma_\mu$ which are called twist eaters. 
Eq.~(\ref{cocycleD}) implies the Weyl-'t~Hooft commutation relations
\beq
\Gamma_\mu \Gamma_\nu
=\ee^{2\pi iQ_{\mu\nu}/p}\,\Gamma_\nu \Gamma_\mu \, .
\label{GammaalgD_higher}
\eeq
In order to find a general solution to~(\ref{bc}), 
we need an explicit form of 
the twist eaters $\Gamma_\mu$ satisfying (\ref{GammaalgD_higher})
for SU$(p)$ and even spacetime dimension $D=2d$. 
For generic rank $p$ and flux matrix $Q$,
these matrices have been constructed in Ref.~\cite{twisteater}
and have the dimension of the irreducible representation of
(\ref{GammaalgD_higher}) equal to
$p/\tilde p_0$ with integer $\tilde p_0$.
The most general gauge field configuration ${U}_\mu(x)$ satisfying
the constraints~(\ref{bc})
is determined by
two integral matrices $\tilde P$ and $B$,
which are constructed in the Appendix, and
can be represented in terms of the operator (\ref{defUhat}) as
\beq
\hat{U}_\mu =
\sum_{\vec{m}}
\,\left(\prod_{\nu=1}^D\left(\hat Z_\nu'\right)^{m_\nu}\right)
~\ee^{  \pi i\sum_{\nu<\lambda}
\Theta ' _{\nu\lambda} \,m_\nu m_\lambda}
\otimes u_\mu(\vec{m})
\ ,
\label{UiZD}
\eeq
where $u_\mu(\vec{m})$ is a $\tilde p_0 \times \tilde p_0$ matrix
and $\hat Z' _\mu$ is given by
\beq
\hat Z'_\mu=\ee^{2\pi i (\Sigma ^{'-1})_{\mu\nu}
\,\hat{x}_\nu}
\otimes\prod_{\rho=1}^D(\Gamma_\rho)^{B_{\mu\rho}} \ ,
\label{Ziansatz_U}
\eeq
where $\Sigma ' = \Sigma \tilde{P}$ and
$\Theta ' = - \tilde{P} ^{-1} B^\top $.
Because of their dependence on the twist eating solutions, 
the operators (\ref{Ziansatz_U}) obey the commutation relations
\beq
\hat{Z}_\mu '\hat{Z}_\nu '
=  \ee ^{- 2 \pi i \Theta ' _{\mu\nu}
}\,\hat{Z}_\nu '
\hat{Z}_\mu '  \ .
\label{Zprimelatticecomm}\eeq
The two matrices $\Sigma '$ and $\Theta '$
must satisfy the general constraint (\ref{MSTdimless}),
which implies that
$M=- \frac{1}{2 \, \epsilon}\,\Sigma   \, B ^\top$ 
must be an integer-valued matrix.
The sum over $\vec{m}$ in (\ref{UiZD})
can then be taken over $\IZ ^D$ modulo the periodicity
$m_\mu \sim m_\mu + \frac1\epsilon\,\Sigma ' _{\nu\mu}$.

We now introduce the map
\beq
\hat{\Delta} ' (x ')=
\frac1{|\det \frac1\epsilon\,\Sigma ' |}\,
\sum_{\vec{m}}
\left(\prod_{\mu=1}^D
\left(\hat Z_\mu'\right)^{m_\mu}\right)~
\ee^{ \pi i \sum_{\mu<\nu}
\Theta ' _{\mu\nu} \,
m_\mu m_\nu} 
~\ee ^{- 2 \pi i
\left(\Sigma ^{' -1}\right) _{\mu\nu}\,m_\mu\,x_\nu'} \,
\label{map_prime}
\eeq
and decompose the operator $\hat{U}_\mu$ as
\beq
\hat{U}_\mu = 
\sum _{x '}\hat{\Delta} ' (x ')\otimes U_\mu ' (x') \ .
\label{newexpansion}
\eeq
It follows that $U_\mu ' (x ')$ is a single-valued 
$\tilde p_0 \times \tilde p_0$ matrix field on a periodic lattice 
with period matrix $\Sigma ' = \Sigma \tilde{P}$ and 
the dimensionless noncommutativity matrix $\Theta '$.
Unitarity of the operator $\hat{U}_\mu$ further requires that
$U_\mu ' (x')$ be star-unitary.
Finally, we arrive at
the action~(\ref{latticeaction}) with reduced rank $n={\tilde p_0}$
and coupling constant $g ^2 = \tilde{g} ^2 (p/{\tilde p_0})$.
This shows that an ordinary U($p$) lattice gauge theory with twisted gauge
fields is equivalent to noncommutative U($\tilde p_0$) lattice gauge theory
with periodic gauge fields.
What we have arrived at is the lattice analog of the well-known fact that, for
$\tilde p_0=1$, noncommutative U(1) Yang-Mills theory with rational-valued
deformation parameters $\Theta _{\mu\nu}$ is 
equivalent to ordinary Yang-Mills
theory with gauge group U$(p)$ and non-vanishing 't~Hooft flux. 
From the present point of view, Morita equivalence is
regarded as a change of basis
$\hat\Delta(x)\leftrightarrow\hat\Delta'(x')$ for 
the mapping between operators
and fields.

We will now discuss the case $\Sigma =\epsilon \id _D$ 
in the present construction. 
A one-site U($p$) lattice gauge theory is just 
the Eguchi-Kawai model \cite{EK},
and the fact that the boundary condition is twisted as in (\ref{bc})
means that it is actually the twisted Eguchi-Kawai
model \cite{GO}.
To see this explicitly, we reduce the action (\ref{commutativeaction})
to a single point $x=0$ by using the constraints (\ref{bc}) and
arrive at the action
\beq
S=-
\frac{1}{\tilde{g}^2}\,
\,\sum_{\mu\neq\nu} {\cal  Z}_{\mu\nu}~\tr_{(p)}\Bigl(V_\mu\,V_\nu\,
V_\mu^\dag\,V_\nu^\dag\Bigr)
\label{EKaction}
\eeq
where $V_\mu=U_\mu (0) \,\Gamma_\mu$, $\mu=1,\dots,D$, are
$p\times p$ unitary matrices. 
This is the action of the twisted Eguchi-Kawai
model, where the phase factor ${\cal Z}_{\mu\nu}$ 
is called the ``twist''. Thus, the recent proposal \cite{AIIKKT} that the
twisted large $N$ reduced model 
serves as a concrete definition of noncommutative
Yang-Mills theory
can be interpreted as the simplest example of Morita equivalence.
The possibility of such an interpretation has also been suggested 
in Ref. \cite{HI}.

\section{Wilson loops on the lattice}
\label{loops}
\setcounter{equation}{0}

We now turn to a description of star-gauge invariant
observables on the lattice~\cite{IIKK,AMNS}. We first define the lattice
parallel transport operator ${\cal U}(x;C)$ which is
specified by a contour
$ C = \{ \mu_1 , \mu_2 , \dots , \mu_n  \}$,
where $\mu_j = \pm 1, \pm 2 , \cdots \pm D$
and $U _{-\mu} (x)= U_\mu (x - \epsilon \hat{\mu})^\dag$.
It is defined by
\beq
{\cal U}(x;C)
= U_{\mu_1} (x) \star U_{\mu _2} (x+ \epsilon \hat{\mu}_1) \star \cdots
\star U_{\mu _n}\left(x+ \epsilon \sum _ {j=1}^{n-1}  \hat{\mu}_j\right) 
\label{lparallel}
\eeq
and it transforms
under the star-gauge transformation (\ref{latticestargaugetr})  as
\beq
{\cal U}(x; C)\mapsto
 g(x)\star {\cal U}(x; C)\star g(x+v )^\dag \ ,
\label{gaugecovariant}
\eeq
where $v=\epsilon\sum_{j=1}^n\hat\mu_j$.

Star-gauge invariant observables are constructed
out of the parallel transport operator ${\cal U}(x;C)$
using a star-unitary function $S_v (x)$ 
with the property 
\beq
S_v(x) \star g(x) \star S_v (x)^\dag= g(x+v)
\label{Svprop}
\eeq
for arbitrary functions $g(x)$ on the periodic lattice. 
The solution to this equation is given by
\beq
S_v (x ) = \ee ^{i k \cdot x }
\label{lSv}
\eeq
provided that
\beq
v_\mu = \theta  _{\mu \nu} k _\nu  + \Sigma _{\mu\nu} n_\nu
\label{vk_relation}
\eeq
with some integer-valued vector $n_\mu$.
Since both $k$ and $v$ are discrete and periodic,
there are the same finite number
of values that $k$ and $v$ can take, by modding out the periodicity.
Therefore, it makes sense to ask whether
(\ref{vk_relation}) gives a one-to-one correspondence between 
$k$ and $v$.
The answer is affirmative if and only if
there exist $D \times D$ integer-valued matrices $J$ and $K$ which satisfy
\beq
\frac{1}{\epsilon}\,\Sigma J - 2 M K =  \id _D \ ,
\label{JKrel}
\eeq
where $M$ is the integer-valued matrix introduced in (\ref{MST}).
If this condition is not met, then there exists $v$ for which 
there is no momentum $k$ satisfying (\ref{vk_relation}), and
for the other $v$ there are more than one 
$k$ satisfying (\ref{vk_relation}).
For example, let us consider the case with 
$\frac{1}{\epsilon}\Sigma = L \id _D$.
If $L$ is even, then (\ref{JKrel}) cannot be satisfied.
If $L$ is odd and $M ^{-1}$ is an integer-valued matrix, then
one can satisfy (\ref{JKrel}) by taking
$J=\id_D$ and $K=\frac{(L-1)}{2} M^{-1}$.

By using the function (\ref{lSv}),
with the property (\ref{Svprop}), 
star-gauge invariant observables
can be constructed out of (\ref{lparallel}) as
\beq
{\cal O} (C) =
\sum_x\tr_{(n)} \Bigl( {\cal U}(x; C) \Bigr)
\star S_v(x) \ .
\label{defcalO}
\eeq
Its star-gauge invariance stems from (\ref{gaugecovariant})
and (\ref{Svprop}).
In Eq. (\ref{defcalO}), 
the parameter $k _\mu$ in 
(\ref{lSv}) can be interpreted as the
total momentum of the contour $C$. 
Note that in the commutative case, $\theta
_{\mu\nu} = 0$, Eq. (\ref{vk_relation}) reproduces the fact that
gauge invariant quantities are given either by closed Wilson loops ($n_\nu=0$)
or by Polyakov loops ($n_\nu \neq 0$), i.e. loops winding around the torus.
Note also that in the commutative case, the total momentum $k _\mu$ is 
unrestricted.
In the noncommutative case, on the other hand, 
one can construct gauge invariant
quantities associated with open loops.
In particular, when the condition (\ref{JKrel}) is met, 
one can construct a gauge invariant quantity for
each contour $C$ on the lattice, and the total momentum
should be specified uniquely (up to periodicity)
through (\ref{vk_relation})
depending on the separation vector $v_\mu$ of the contour.
We will see in the next
Section that star-gauge invariant observables constructed
in noncommutative Yang-Mills theory reduce smoothly to 
ordinary Wilson loops in the commutative limit.

\section{Coupling to fundamental matter fields}
\label{hopping}
\setcounter{equation}{0}

We will now consider noncommutative gauge theories coupled to matter
fields in the fundamental representation of the gauge group.
One advantage of the lattice formulation is that it allows
a hopping parameter expansion (or large mass expansion)
of this theory which will enable us to clarify
various aspects of the star-gauge invariant observables  constructed
 out of noncommutative gauge fields. 
For simplicity, we consider a complex scalar field $\phi(x)$
as the matter field. Fermions can be treated in an analogous way.
The action for the matter field is
\beq
S_{\rm matter}
= - \kappa   \left\{
\sum _{x, \mu} \phi ^\dag (x) \star U_\mu (x) \star
\phi (x + \epsilon \hat{\mu})
+ \mbox{c.c.} \right\}
+ \sum _x \phi ^\dag (x)\,\phi (x) \ ,
\label{matter_action}
\eeq
and it is invariant under the star-gauge transformation
\beq
\phi (x) \mapsto  g(x) \star \phi (x) ~~~;~~~
\phi ^\dag (x)  \mapsto  \phi ^\dag (x) \star g^ \dag (x)   ~~~;~~~
U_\mu (x) \mapsto  g (x) \star U_\mu (x) \star
g ^\dag  (x + \epsilon \hat{\mu}) \ .
\label{gaugetr_matter}
\eeq
By integrating over the field $\phi (x)$, we perform an expansion
in the hopping parameter $\kappa$.

The effective action $\Gamma _{\rm eff} [U]$
for the gauge field $U_\mu (x)$  induced
by the integration over $\phi (x)$ 
is  given by
\beq
\Gamma _{\rm eff} [U] = - \ln \sum_{n=0}^\infty\frac{\kappa ^n}{n ! }\,
\left\langle
\left( 
\sum _{x, \mu} \phi ^\dag (x) \star U_\mu (x) \star
\phi (x + \epsilon \hat{\mu})+ \mbox{c.c.} \right)^n
\right\rangle _{\kappa = 0} \ .
\eeq
Throughout this section, $\langle \cdots \rangle$ refers to
the vacuum expectation value for fixed gauge background,
namely we integrate over the matter fields only.
Using Wick's theorem, we obtain%
\footnote{A calculation of the hopping parameter expansion 
with fundamental fermions
has been performed by A.~Zubkov.}
\beq
\Gamma _{\rm eff}[U] = \sum _C \frac{\kappa ^{L(C)}}{L(C)}\,\sum _x
\tr_{(n)}\Bigl( {\cal U}(x;C)\Bigr) \ ,
\label{inducedaction}
\eeq
where $\sum _C$ denotes the sum over all closed loops 
on the lattice and $L(C)$ denotes the length
of the loop $C$ in lattice units.
In this way, we encounter observables associated 
with closed loops\footnote{We can also introduce $n_f$ flavours of the matter
field and take the limit $n_f\rightarrow\infty$ with 
$\kappa\sim n_f^{-1/4}$. In that case, only the
simplest closed loops of lengths 4, i.e. the 
boundaries of lattice plaquettes,
remain in (\ref{inducedaction}). The single-plaquette 
action~(\ref{latticeaction})
can thus be induced by fundamental matter fields, similarly to
the commutative case~\cite{indu}.}.

Let us now consider star-gauge invariant observables involving
matter fields such as
\beq
G[f] = \left\langle \sum_ x\phi ^\dag (x) \star \phi (x)
\star f(x)\right\rangle \ ,
\eeq
which is star-gauge invariant for arbitrary functions $f(x)$ on the lattice
which can be regarded as the wavefunction of the composite operator
$\phi ^\dag (x) \star \phi (x)$.
We first make a Fourier transform of $f(x)$ and express $G[f]$ as
\beq
G[f] = \sum_{\vec k}\tilde{f} (k)
\left\langle
\sum_x\phi ^\dag (x) \star \ee ^{i k\cdot x} \star \phi (x+v)
\right\rangle \ ,
\eeq
where $v_\mu = \theta _{\mu\nu} k_\nu $.
We have used the fact that $\ee ^{i k\cdot x}$ acts as a
translation operator in the noncommutative field theory as follows from
Eqs.~(\ref{Svprop}) and (\ref{lSv}).
Integrating over the matter fields using the hopping parameter expansion,
we obtain 
\beq
G[f] = \sum _{\vec k} \tilde{f}(k)\,\sum _C \kappa ^{L(C)}\,\sum _x
\tr_{(n)}\Bigl( {\cal U}(x;C) \star \ee ^{i k\cdot x }\Bigr)
\eeq
where $\sum _C$ now denotes the sum over all lattice loops
beginning at the origin and ending at $v_\mu$.
Thus we find the observables encountered in the previous Section
which are associated with open loops. In the commutative case,
the separation vector $v_\mu$ of the loops
should vanish independently of the momentum $k_\mu$.
Thus the summation over the loop $C$ contains only closed loops.
The summation over $\vec k$ can then be done explicitly
reproducing the wavefunction $f(x)$.
In the noncommutative case, the summation over the loop $C$ depends
on $k_\mu$ and we are not allowed to reverse the order of the summations.
The larger $k_\mu$ is, the larger becomes the separation vector $v_\mu$
of the two ends of the loop $C$.
This is a characteristic phenomenon in noncommutative field theories.
If one would like to have a higher resolution in one direction,
say in the $\mu = 1$ direction, by increasing $k_1$, then
the object will extend
in the other directions proportionally to $\theta _{\mu 1} k_1$.
On the other hand, if one considers the case when $\tilde{f}(k)$ has support
only for finite momentum $k_\mu$, then
one can take the $\theta _{\mu\nu} \rightarrow 0$ limit smoothly,
reproducing the commutative case.

The considerations in this Section show that the star-gauge
invariant observables 
indeed play the same fundamental role as ordinary Wilson loops do
in commutative gauge theories.
As the knowledge of all the Wilson loop correlators would give
all the information of the gauge theory including matter fields,
so do the star-gauge invariant observables in the noncommutative case.
We have also seen explicitly how these observables reduce smoothly to
ordinary Wilson loops in the commutative limit.

\section{Morita equivalence with fundamental matter fields}
\label{Morita_fundamental}
\setcounter{equation}{0}

In this final Section, we discuss
Morita equivalence of noncommutative Yang-Mills theories coupled to
fundamental matter fields.
We show, in particular, that the twisted Eguchi-Kawai model
with fundamental matter as constructed in Ref.~\cite{Das} 
is equivalent to noncommutative Yang-Mills theory with 
fundamental matter fields. We use the setup of Section~\ref{lattice}.
We consider {\em commutative} U($p$) gauge theory and 
introduce $N_f$ flavours of matter in the fundamental representation.
We represent it as $\Phi (x) _{ij}$, where $i=1,\dots, p $ represents
the color index and $j = 1,\dots , N_f$ represents the flavour index.
The spacetime is discretized as $x_\mu\in\epsilon \IZ$.
For the present purpose, it is essential to take $N_f$ to be
an integral multiple of $p$.
In what follows, we assume that $N_f = p$ for simplicity.
Then, $\Phi (x) _{ij}$ is a $p \times p$ general complex matrix.

The action for the matter part of the U($p$) gauge theory is given by
\beq
S_{\rm matter}
= - \kappa
\left\{ \sum _{x, \mu} \tr_{(p)}\Bigl( \Phi ^\dag (x)  U_\mu (x)
\Phi (x + \epsilon \hat{\mu})\Bigr) + \mbox{c.c.} \right\}
 + \sum _x \tr_{(p)}\Bigl(\Phi ^\dag (x)  \Phi (x)\Bigr) \ ,
\label{matter_action_Morita}
\eeq
which is invariant under 
the gauge transformation 
$U_\mu (x) \mapsto g(x) U_\mu (x) g(x+\epsilon \hat{\mu})$ and
$\Phi (x) \mapsto  g(x)  {\Phi} (x)$.
The action (\ref{matter_action_Morita}) has also
a global U($N_f$) flavour symmetry
$\Phi (x) \mapsto  {\Phi} (x)\,g'$,
where $g'\in \mbox{U}(N_f)$.
We impose a constraint on $\Phi(x)$ and $U_\mu(x)$ given by the twisted
boundary conditions
\beq
U_\mu  (x+ \Sigma _{\alpha\nu}\hat{\alpha}) =
\Gamma _\nu\,U_\mu  (x)\,\Gamma _\nu  ^\dag \ , ~~~~~
\Phi  (x+ \Sigma _{\alpha\nu}\hat{\alpha}) =
\Gamma _\nu\,\Phi  (x)\,\Gamma _\nu  ^\dag \ ,
\label{Phi_constraint}
\eeq
where $\Sigma$ is the period matrix.
Since the transition matrix lives in SU$(p)$, this implies a
nonvanishing 't~Hooft flux of the gauge field, as we discussed in
Section~\ref{lattice}.
Notice also that 
in (\ref{Phi_constraint}), the $\Gamma _\nu$ on the left of $\Phi (x)$ 
represents a (global) gauge transformation, whereas the 
$\Gamma _\nu ^\dag$ on the right of $\Phi (x)$ represents
a rotation in the flavour space.
This is the trick to introducing the fundamental representation in
Morita equivalence.
We have made use of the global flavour symmetry SU($N_f$) to mimic
the boundary condition for the adjoint representation.
Such an idea first appeared in Ref.~\cite{CG}
in the context of supersymmetric theories.

Solving the constraint as we did in Section \ref{lattice},
we find that the resulting theory is 
{\em noncommutative} U($\tilde{p}_0$) lattice gauge theory
coupled to $\tilde{p}_0$ flavours of matter fields in the fundamental
representation, with periodic boundary conditions.
We can set $\Sigma=\epsilon \id _D$ 
to obtain the twisted Eguchi-Kawai model
with matter fields.
This is the model introduced in Ref.~\cite{Das} as a model 
which reproduces large $N$ gauge theory with
$N_f$ flavours in the Veneziano limit $N_f \sim N \rightarrow \infty$.
Now, we have found that the same model with
$N_f = n_f N$ can be interpreted as 
noncommutative U($\tilde{p}_0$) gauge theory
with periodic boundary conditions including 
$n_f \tilde{p}_0$ flavours of matter fields in the fundamental representation.
The one-loop beta-function for $\tilde{p}_0 = 1$ 
and fermionic matter has been calculated in Ref.~\cite{Hayakawa}.
The result agrees with
SU($N$) commutative Yang-Mills theory with $N_f = n_f N$ flavours
as it should due to Morita equivalence. 
In the present case, there is no extra
infrared singularity associated with non-planar diagrams~\cite{uvir}
since one-loop diagrams with fundamental matter are always planar. 

\subsection*{Acknowledgements}

The work of J.A. and Y.M. is supported in part by
MaPhySto founded by the Danish National Research Foundation.
Y.M. is sponsored in part by the Danish National Bank.
J.N. is supported by the Japan Society for the Promotion of
Science as a Research Fellow Abroad. 
The work of R.J.S. is supported in part by
the Danish Natural Science Research Council.


\setcounter{section}{0}
\appendix{General representation of twist eaters}
\label{append}
\setcounter{equation}{0}

The explicit form of 
the twist eaters $\Gamma_\mu$ satisfying (\ref{GammaalgD_higher})
for SU$(p)$ with generic rank $p$, flux matrix $Q$ and even dimension $D=2d$
can be constructed as follows~\cite{twisteater}.
Using the discrete symmetry of the $D$-dimensional torus under the
SL$(D,\IZ)$ geometrical automorphism group, $Q$ can be represented 
in a canonical skew-diagonal form
\beq
Q=\pmatrix{0&-q_1& & & \cr q_1&0& & & \cr & &\ddots& & \cr & & &0&-q_d\cr & &
&q_d&0\cr} \ .
\label{Qdiag}\eeq
Given the $d$ independent fluxes $q_i\in\IZ$, we introduce the integers
$p_i={\rm gcd}(q_i,p)$, $\tilde p_i= p/{p_i}$, and $\tilde q_i={q_i}/{p_i}$.
By construction, $\tilde p _i$ and $\tilde q _i$ are co-prime.
A necessary and sufficient condition for the existence of solutions to
(\ref{GammaalgD_higher}) is
that the integer $\tilde p_1\cdots\tilde p_d$,
the dimension of irreducible representation of the Weyl-'t~Hooft algebra,
divides the rank $p$ \cite{twisteater}. In that case we write
\beq
p=\tilde p_0\,\prod_{i=1}^d\tilde p_i
\label{tildep0def}\eeq
and the twist eating solutions may then be given on the subgroup ${\rm
SU}(\tilde p_1)\otimes\cdots\otimes{\rm SU}(\tilde p_d)\otimes{\rm SU}(\tilde
p_0)$ of SU$(p)$ such that $\Gamma_{i-1}, \Gamma_i $ are constructed
from the Weyl-'t~Hooft matrices on ${\rm SU}(\tilde p_i)$.
The subgroup of GL$(p,\IC)$
consisting of matrices which commute with the twist eaters $\Gamma_\mu$ is then
GL$(\tilde p_0 , \IC)$. 

The general solution to Eq.~(\ref{bc}) for ${U} _\mu (x)$ is determined by
two $D\times D$ integral matrices which,
in the basis (\ref{Qdiag}) where $Q$ is skew-diagonal, read
\beq
\tilde P=\pmatrix{\tilde p_1&0& & &
\cr0&\tilde p_1& & & \cr & &\ddots& & \cr & & &\tilde p_d&0\cr & & &0&\tilde
p_d\cr}~~~~~~,~~~~~~B=\pmatrix{0&-b_1& & & \cr b_1&0& & & \cr & &\ddots& &
\cr & & &0&-b_d\cr & & &b_d&0\cr}
\label{tildeQPdiag}\eeq
with integral  $b_i$ obeying
$a_i \tilde p_i+ b_i \tilde q_i =1$, $i=1,\dots,d$,
for some integer $a_i$. 
Defining matrices $A$ and $\tilde{Q}$ using $a_i$ and $\tilde{q}_i$
as with $\tilde{P}$ and $B$ in (\ref{tildeQPdiag}), respectively,
the previous condition can be written as $A\tilde P+B\tilde Q=\id_D$,
which is invariant under the SL$(D,\IZ)$ transformation
\beqa
&~&A \mapsto \Lambda ^{\top} \, A \,  \Lambda 
~~~~~~~~~~~~~~;~~~~~
\tilde{P} \mapsto \Lambda ^{-1} \,  \tilde{P} \,(\Lambda^{-1})^\top
\n
&~&B \mapsto \Lambda ^{\top} \, B \, \left(\Lambda ^{ -1}\right)^\top 
~~~~~;~~~~~
\tilde{Q} \mapsto \Lambda ^\top \, \tilde{Q} \,(\Lambda^{-1})^\top \ ,
\eeqa
where $\Lambda \in{\rm SL}\,(D,\IZ)$.
The flux matrix $Q$
can be written in terms of $\tilde{P}$ and $\tilde{Q}$
in an SL$(D,\IZ)$ covariant way
as $Q=p \,\tilde{Q}\, \tilde{P}^{-1}$, and it transforms as 
$Q \mapsto \Lambda ^\top \, Q \, \Lambda $.
One can use this symmetry to rotate $Q$ back to general form and thus
 find the corresponding
$\tilde{P}$ and $B$ which are needed to construct the general solution
to Eq.~(\ref{bc}).

\end{document}